\documentstyle[prl,multicol,aps,epsf]{revtex}
\renewcommand{\narrowtext}{\begin{multicols}{2} \global\columnwidth20.5pc}
  \renewcommand{\widetext}{\end{multicols} \global\columnwidth42.5pc}

\begin{document}
\draft
\title{Kondo Temperature for the Two-Channel Kondo Models of Tunneling Centers
}
\author{I. L. Aleiner$^{1,2}$, B. L. Altshuler$^{2,3,4}$,  
Y. M. Galperin$^{2,5}$, and T. A. Shutenko$^{2,4}$  }
\address{$^1$Department of Physics and Astronomy,
SUNY at Stony Brook, Stony Brook, NY 11794, USA;\\
$^2$Centre for Advanced Studies, Drammensveien 78, Oslo, Norway;\\
$^3$NEC Research Institute, 4 Independence Way, Princeton, NJ 08540, USA;\\
$^4$Physics Department, Princeton University, Princeton, NJ 08544, USA;\\
$^5$Physics Department, University of Oslo, PO Box 1048 Blindern, 0316
Oslo, Norway and Division of Condensed Matter Physics, A. F. Ioffe
Institute, 104021 St. Petersburg, Russia.}

\maketitle
\begin{abstract}
The possibility for a two-channel Kondo  ($2CK$) non Fermi liquid state
to appear in a metal as a result 
of the  interaction between electrons and 
movable structural defects is revisited.
As usual, the defect is modeled by a heavy particle moving 
in an almost symmetric double-well potential (DWP). 
Taking into account only the two lowest states in DWP 
is known to lead to a Kondo-like Hamiltonian 
with rather low Kondo temperature, $T_K$. 
We prove that, in contrast to previous believes, 
the contribution of higher excited states in DWP does not enhance $T_K$. 
On the contrary, $T_K$ is reduced by three orders of magnitude 
as compared with the two-level model: 
the prefactor in $T_K$ is determined 
by the spacing between the second and the third levels in DWP
rather  than by the electron Fermi energy. 
Moreover, $T_K$, turns out to be parametrically smaller than 
the splitting between the two lowest levels. 
Therefore, there is no microscopic model 
of  movable defects which may justify non-Fermi liquid $2CK$ phenomenology.
						
\end{abstract}

\narrowtext
It is well known that two-level systems (TLS)
determine the low energy phenomena in a glassy matter.
The most popular realization of the TLS
is a movable atom 
tunneling between two minima
of the two-well potential created by other atoms\cite{AHV}. 
The low-temperature behavior
of glasses was found to be consistent with the assumption
of homogeneous distribution of both energy difference and
spatial distance $a$ between the minima.   
In metallic glasses TLS interact with itinerant
electrons. 
Usually this interaction in metallic glasses is assumed to be weak 
and to manifest itself only in a finite relaxation rate 
of the TLS, see Ref.~\onlinecite{Black} for a review. 

It was proposed long ago ~\onlinecite{StromOlson,Kondo76,Zawadowski80} 
that a TLS interacting with itinerant electrons 
behaves like
a localized spin in the Kondo model.  
Indeed, in the limit 
$k_Fa\ll 1$, where $k_F$ is the Fermi wavelength,
only the electrons with two spherical harmonics, namely, 
$l=0$ and $l=1,m=0$ interact with TLS. 
(Here and below the axis of the momentum quantization is 
the easy axis of TLS, $x$)

Let us introduce a pseudospin $\hat S$ of a symmetric TLS: 
$S=-1/2$ corresponds to the ground state
(even wave function),  
whereas $S=1/2$ labels the excited state with the odd wave function.
One can map the electrons with the two relevant spherical harmonics 
on the one-dimensional Fermi gas of particles that
are characterized by a {\em pseudo-spin} 
with components $\sigma=+,- $ as
$\hat{\Psi}_{l=0}\equiv \hat{\Psi}_{-}$, 
$\hat{\Psi}_{l=1,m=0}\equiv \hat{\Psi}_{+}$,
while the real electron spin index $s=\uparrow , \downarrow$ 
is replaced with the channel index $\mu=0,1$. 
Furthermore, provided that 
the Fermi energy, $\varepsilon_F$, exceeds
all of the relevant energies,
one can linearize the electron dispersion law
near the Fermi level, $\varepsilon (p)\simeq v_F|p|$,
where $v_F$ is the Fermi velocity. 
The divergences caused by the linearized spectrum should, 
thus, be cut off by the bandwidth $D\simeq \varepsilon_F$.
The resulting Hamiltonian of the system can be
expressed as
\begin{eqnarray}
&&\hat{H}  =-iv_F\int_{-\infty}^{\infty}\! dx\sum_{\mu=0,1}\,
\sum_{\sigma=\pm } \Psi^{\dagger}_{\mu,\sigma}
\partial_{x}\Psi_{\mu,\sigma}+ \sum_{i =x,z}\Delta_i \hat{S}_i 
\nonumber\\&& \quad
+2\pi v_F\sum_{j=x,y,z}\,\sum_{\mu=0,1}\,
\sum_{\sigma,\sigma^{\prime}=\pm }
v_j \Psi^{\dagger}_{\mu,\sigma}\hat{\tau}^j_{\sigma\sigma^{\prime}}
\Psi_{\mu,\sigma^{\prime}}\hat{S}_j\, .
\label{eq:1.1}
\end{eqnarray}
Here the Pauli matrices, $\tau^j_{\sigma\sigma^{\prime}}$,
act in the space of the electron pseudospin, and
$\hat{S}$ is the operator of the TLS pseudospin, 
$\left[S^i,S^j\right]=i\epsilon^{ijk}S_k$. 
The first term in Eq.~(\ref{eq:1.1}) 
describes kinetic energies of 1D electrons. 
The second term characterizes the TLS level splitting:
$\Delta_z$ and $\Delta_x$ represent correspondingly the tunneling 
and the initial TLS asymmetry. 
The third term in Eq.~(\ref{eq:1.1}) describes
TLS-electron interaction.
The Hamiltonian Eq.~(\ref{eq:1.1}) is nothing but the two-channel
Kondo Hamiltonian\cite{NB}, where the level splitting plays the role
of the Zeeman splitting of states of the usual Kondo impurity. 
For a  comprehensive review of implication of this model to
magnetic ions and tunneling centers in metals 
see Ref.~\onlinecite{CoxZawadowski}. 

The two-channel Kondo effect is known to manifest itself 
through a non-Fermi liquid behavior of the specific heat,
magnetization and electronic correlation functions. 
Such a behavior takes place when
both the temperature, $T$, and the level splitting,
\begin{equation}
\Delta=\sqrt{\Delta_x^2+\Delta_z^2}\, , 
\label{Delta}
\end{equation} 
do not exceed the Kondo temperature, $T_K$. 
It can be shown\cite{VZ,VZ1} that in the limit $v_z\ll v_x\ll 1$
\begin{equation}
T_K=D\left(v_x v_z\right)^{1/2}
\left(v_z \over 4v_x\right)^{1/4v_x}\, .
\label{eq:1.2}
\end{equation}
The non-Fermi liquid behavior of the TLS at the
two-channel Kondo fixed point was used in 
Ref.~\onlinecite{RLDB} to interpret the 
zero bias anomaly in characteristics of point
 contacts \onlinecite{KR} and more recently \onlinecite{RDZ}
 to explain the temperature behavior of the dephasing
 rate observed in Refs.~\onlinecite{MW,Pothier}. In general, TLS
 were assumed to play important role in crystalline metals 
as well as in  metallic glasses. 

Assumptions of Ref.~\onlinecite{RDZ} were criticized on the grounds
that the disorder induced splitting estimates were too high for the
Kondo-like behavior to develop \onlinecite{RefD}.  Although several
questions raised in Refs.~\onlinecite{RefD} remain unanswered, we put
this issue aside.  Instead, we concentrate on a different objection -
smallness of $T_K$ in Eq.~(\ref{eq:1.2}) for reasonable values of
parameters.  Here we prove that $T_K \ll \Delta_z$ for any set of the
microscopic parameters, which allows the Kondo like description,
Eq.~(\ref{eq:1.1}).  Therefore, the two channel Kondo fixed point by
no means can  be reached with the lowering $T$ and thus is
irrelevant for the description of the TLS in metals.

To understand why the resulting Kondo temperature is so small,  
let us first discuss the physical meaning of
the bare coupling  constants $v_j$ and estimate them. 

The coupling constant $v_x$ in  Eq.~(\ref{eq:1.1}) determines
the renormalization of the TLS asymmetry 
by the electrons - a tilting of the double-well potential
by the dipole moment of the electron density.  
Assuming a contact interaction characterized by a 
dimensionless coupling constant $\lambda <1$ 
we can estimate $v_x$
at given Fermi wave number, $k_F$, and the size of TLS, $a$,
as (see, e.g., ~\cite{Black,CoxZawadowski})
$v_x\simeq \lambda k_Fa$.

As to $v_z$, it characterizes transition between the 
two states of the TLS assisted by an electron transition. 
Incoming electron renormalizes the barrier's height, 
$V$, and consequently the tunneling amplitude. 
However, the tunneling event still has to occur.
Therefore, ~\cite{Black,CoxZawadowski} 
$v_z\simeq \lambda k_F^2a^2 \exp (-\eta)$, 
where the tunneling exponent $\eta $  
is determined by $V$ and the atomic mass $M$.  
Since $M$ is large, $\eta \gg 1$ even for relatively low barriers. 
As a result, the coupling constant $v_z$ is always much smaller than $v_x$:
\begin{equation}
v_z\simeq v_x(k_Fa)\exp (-\eta )\ll v_x\, ,\ \ 
\eta \simeq \hbar^{-1}a\sqrt{8MV} 
\label{vsmall}
\end{equation}
This is why $v_z$ was usually neglected in previous treatments of TLS.  
For ``typical'' values of the parameters 
\begin{equation}
D\sim 5 \mbox{eV},\quad
 v_x\simeq 0.2, \quad v_z/v_x\simeq 10^{-3}, 
\label{typical}
\end{equation} 
the ``conventional'' estimate of the Kondo temperature is 
$T_K^{c}\simeq 10^{-2}-10^{-3}$ K \cite{VZ1}. This low value of $T_K$   
makes it hard to believe that the Kondo fixed point  
is relevant for the discussion of existing experiments.

In attempt to resolve the problem of small $T_K$ 
authors of Ref.~\cite{ZarandZawadowski}   
went beyond the two level approximation and 
considered virtual tunneling through the third level of the ``TLS''. 
This contribution to $v_z$ apparently does not contain
the tunneling exponent.
According to Ref.~\onlinecite{ZarandZawadowski},  
this fact dramatically increases  $T_K$ comparing to the 
``conventional'' estimate. 
In our opinion the statement about the large 
increase in the Kondo temperature is \emph{incorrect}.
Below we discuss the problem in detail. 
  
$T_K$ can be extracted from  second order perturbation
theory in the interaction of the tunneling particle
with the  electrons.
We calculate this correction  
within the one-dimensional (1D) model suggested in
Ref.~\onlinecite{ZarandZawadowski}.   

Consider a heavy particle in a symmetric 1D 
double-well potential $V(x)$.
Let the energies of the two lowest eigenstates, $E_{1,2}$,
be indistinguishable: 
$E_2-E_1 \to 0$. 
One can express the matrix element 
of the contact interaction of this particle with electrons 
through the coupling constant, $\lambda $, Fermi velocity, $v_F$,
the heavy-particle eigenfunctions, $\phi_i(x)$,
and the electron wave functions, 
$\psi_{\sigma}(x)$, with a given isospin, $\sigma=+,-$:  
\begin{equation}
U^{\sigma_1\sigma_2}_{ij} = 2\pi v_F\lambda I^{\sigma_1\sigma_2}_{ij}\, , \ \
I^{\sigma\rho}_{kl}\equiv \int dx \,
\phi_k \phi_l \psi_{\sigma}
\psi_{\rho} \, .
\label{matrix_element}
\label{eq:5.1}
\end{equation}

There are two second order corrections
to the scattering amplitude ${\sigma_1,i} \to {\sigma_2,j}$, 
which correspond to  processes with different intermediate states:

(i) In the intermediate state an electron  
has an isospin $\sigma$ and the particle  
occupies a state $k$ 
with the energy $E_k\equiv E_1+\varepsilon_k$
(see Fig.~\ref{Fig2}a).
{\begin{figure}[ht]
\epsfxsize=4cm
\centerline{\epsfbox{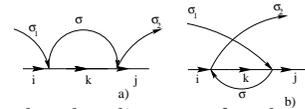}}
\caption{Second order diagrams for the scattering amplitude.
a) electron process. b) hole process.
}
\label{Fig2}
\end{figure} 
}
(ii)
The transition $i\to k$ of the heavy particle 
produces an electron-hole pair - an electron in the final state
$\sigma_2$ and a hole in the state $\sigma$.
Afterwards the hole annihilates the electron in the initial
state $\sigma_1$ and the heavy particle changes its state
from  $k$ to the final state $j$ (Fig.~\ref{Fig2}b)

Combining the contributions of these two processes 
and taking into account the occupation numbers 
of electron states at a given temperature $T$ 
we present the second order correction to the matrix element as:
\begin{equation}
\delta U^{\sigma_1\sigma_2}_{ij}\propto \lambda^2v_F\sum_{k,\sigma}
\int^D_{-D}\! \frac{d\xi \,
\left( I_{ik}^{\sigma_1\sigma}I_{jk}^{\sigma_2\sigma}-
I_{jk}^{\sigma_1\sigma}I_{ik}^{\sigma_2\sigma} \right) }
{\left( -\varepsilon_k-\xi \right) 
\left[ 1+\exp \left( -\xi /T\right) \right] }     \, . 
 \label{eq:5.2} \\
\end{equation} 
Here $I_{kl}^{\sigma \rho}$ is determined by Eq.~(\ref{eq:5.1}),
$\xi$ denotes the energy of the 
intermediate electron or hole state 
counted from the Fermi level, and
$\varepsilon_k\equiv E_k-E_1$ is the energy of the $k$-th state
in the double-well potential
counted from its ground state $E_1=E_2$.
We have already discussed that the domain of the integration 
over $\xi $ should be $|\xi |<D \sim \varepsilon_F$.
As to the singularity at $\xi = -\varepsilon_k$,  
the integral in Eq.~(\ref{eq:5.2}) should be understood as the
principal value.
The minus sign in front of the second (hole) term in the 
numerator is due to the anti-commutation of fermionic operators.

Consider now matrix elements 
$U_{12}^{\sigma_1\sigma_2 }=-U_{21}^{\sigma_1\sigma_2 }$ 
that describe transitions 
of the heavy particle between its two lowest states.
Using  Eqs.~(\ref{eq:5.2}) one can show that 
such a transition should be accompanied by the
change of the electron isospin: due to the parities of the wave 
functions $U_{12}^{--}=U_{12}^{++}=0$. 
To evaluate $U_{12}^{-+}$ we
sum over $\sigma $ and integrate over $\xi $ in  Eqs.~(\ref{eq:5.2}).
The result can be written as
\begin{eqnarray} 
\frac{\delta U^{-+}_{12}}{U^{-+}_{12}}&=&\lambda u_{\infty}\, , \ \ 
u_n\equiv \frac{1}{I_{12}^{-+}}\sum_{k=1}^{n}
\sum_{i,j=1}^2\epsilon^{ij}c_{ij}^{k} 
\ln \left(\frac {D}{\varepsilon_k^{*}}\right) \, , 
\label{eq:5.5}
\end {eqnarray}
where $\epsilon^{11}=\epsilon^{22}=0$,  
$\epsilon^{12}=-\epsilon^{21}=1$, and
\begin{eqnarray}
&&c_{ij}^k\equiv I_{ik}^{+-}(I_{jk}^{--}-I_{jk}^{++}) \,   , \ \
\varepsilon_k^{*} \equiv \max \left \{ \varepsilon_k,T \right \} \, . 
\label{eq:5.5a}
\end{eqnarray}
We start with the contribution of the first two levels ($k=1,2$) to the sum
over $k$ in Eq.~(\ref{eq:5.5}). 
Using Eqs.~(\ref{eq:5.2}), and
(\ref{eq:5.5a}) one obtains, cf. with
Ref.~\onlinecite{ZarandZawadowski},
\begin{equation}
u_2=\ln\left(\frac{D}{T}\right)
\int dx \, \left[\phi_2^2-\phi_1^2\right]
\left[\psi_{-}^2-\psi_{+}^2\right]\, ,
\label{eq:6.1}
\end{equation}
The electron wave functions $\psi_{\sigma}(x)$ are standing waves,
$\psi_{-}(x)+i\psi_{+}(x)=\sqrt{2}\exp (ik_Fx)$.
Since $k_Fa\ll 1$ (otherwise this 1D approach is not applicable)
\begin{equation}
\psi_{-}(x)\simeq \sqrt{2}\left[1-(k_Fx)^2/2\right]\, , \ \
\psi_{+}(x)\simeq \sqrt{2}k_Fx\, . 
\label{psi}
\end{equation}
We introduce the wave functions, 
$\phi_{l(r)}$, localized in the left (right) well,
$\sqrt{2}\phi_{l,r}=\phi_1\pm \phi_2$, recall that
the functions $\phi_i$ are normalized, 
and rewrite Eq.~(\ref{eq:6.1}) as
\begin{equation}
u_2=8\ln \left( \frac{D}{T}\right)
\int dx \, \left(k_Fx\right)^2\phi_l(x)\phi_r(x)\, .
\label{finu}
\end{equation}
The wave functions $\phi_l(x)$ and $\phi_r(x)$ are localized in
the different wells, their overlap being exponentially small.
Accordingly, $u_2$ is exponentially small as well,
and one arrives at the ``conventional'' estimate for $T_K$.

It turns out that taking into account the higher excited states
in the double-well potential, i.e., terms with $k>2$ 
in the sum Eq.~(\ref{eq:5.5}), can only reduce the estimation of $T_K$.

Indeed, it follows from the definition of $I_{ij}^{\sigma \rho}$, 
Eq.~(\ref{eq:5.1}), and completeness of the set 
of the functions, $\{ \phi_k\}$,
($\sum_k \phi_k(x)\phi_k(y)=\delta (x-y)$) that the sums, 
$\sum_k I_{ik}^{\sigma_1\sigma_2}I_{jk}^{\sigma_3\sigma_4}$
with any set ${\sigma_n}$ and consequently $\sum_kc_{ij}^k$, 
are symmetric with respect to the permutation of $i$ and $j$, i.e.,
$\sum_kc_{ij}^k=\sum_kc_{ji}^k$. As a result 
\begin{equation} 
w_{\infty}=0\, ,\ \ \
w_n\equiv \sum_{k=1}^n\sum_{i,j=1}^2\epsilon^{ij}c_{ij}^k\, .
\label{sumrule}
\end{equation} 
The sum rule, Eq.~(\ref{sumrule}), together with Eq.~(\ref{eq:5.5}) imply 
that the second order correction Eq.~(\ref{eq:5.5a}) 
to the matrix element $U_{12}^{-+}$ {\em vanishes at high
temperatures}, when $\varepsilon_k^{*}=T$.
More precisely, the usual Kondo logarithmic temperature dependence of 
$U_{12}^{-+}$ persists only for $T<\varepsilon_3\equiv E_3-E_1$.
Therefore, in the expression for $T_K$, 
Eq.~(\ref{eq:1.2}), the bandwidth 
$D\sim \varepsilon_F\sim $5eV should be substituted by
$\varepsilon_3\sim $3meV:

\begin{equation}
T_K=\varepsilon_3\left(v_x v_z\right)^{1/2}
\left(v_z \over 4v_x\right)^{1/4v_x}\, .
\label{TKcorrect}
\end{equation}
It means that {\em the Kondo temperature is about
three orders of magnitude less than the ``conventional'' estimate!}

To interpret this result note that when it tunnels, the heavy particle
is under the barrier for a time $\sim \hbar/\varepsilon_3$.
At energies bigger than $\varepsilon_3$ we thus deal 
with a continuously moving particle rather than a pseudospin. 
 
However, any truncation of the sum, Eq.~(\ref{eq:5.5}), 
results in a strong overestimation of the Kondo temperature. 
Indeed, the truncated sum $w_{n<\infty }$, Eq.~({\ref{sumrule}), 
is neither zero, nor exponentially small. 
Thus $u_n=w_n \ln (D/T)\gg u_2$ at least when $\varepsilon_n\ll T$.
It is the substitution of $u_{\infty }$ by $u_3$, 
proposed in Ref.~ \onlinecite{ZarandZawadowski}, 
that dramatically enhanced $T_K$.

The sum rule Eq.~(\ref{sumrule}) hints that 
although contribution of each excited state $k$ 
to $u_n$ is quite large in absolute value (if $k$ is not too big), 
these contributions have different signs 
and cancel each other up to an exponentially small quantity 
$u_{\infty}$ when {\em all of them} are included.

To demonstrate that this is the case 
we repeated numerical calculations of 
Ref.~\onlinecite{ZarandZawadowski}, using the same model potential,
Fig.~\ref{Fig3},  but took
into account all of the excited states $\{k\}$ rather than only $k=3$.

Following Ref.~\onlinecite{ZarandZawadowski} 
we chose the barrier height to be 
$V=9.86\hbar^2/2Mb^2$, 
where $M$ is the particle mass 
and b is the well width. 
We computed the eigenfunctions $\phi_i(x)$ and
used  Eqs.~( \ref{eq:5.5}),( \ref{eq:5.5a}) and (\ref{eq:5.1}) to
evaluate $\tilde{u}_n\equiv u_n(k_Fb)^{-2}$.
\begin{figure}
\epsfxsize=3cm
\centerline{\epsfbox{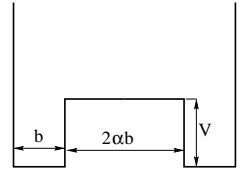}}
\vspace*{0.2cm}
\caption{ Symmetric double well potential with the well width $b$,
the barrier has a hight $V$ and a width $2\alpha b$.
}
\label{Fig3}
\end{figure} 
}

\begin{figure}[ht]
\epsfxsize=8cm
\centerline{\epsfbox{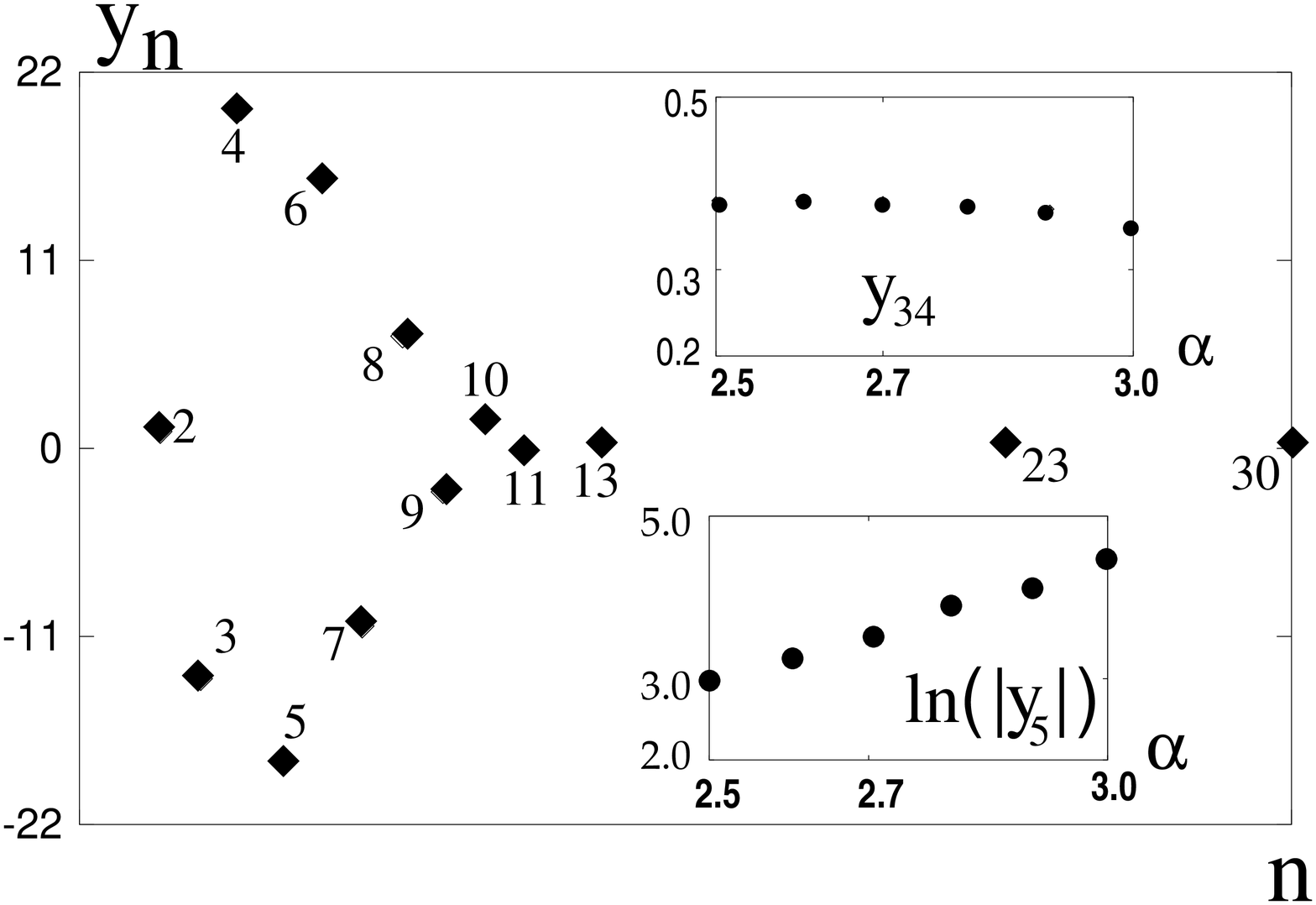}}
\vspace*{0.2cm}
\caption{$y_n\equiv u_n/u_2$ as a function of the level number n
and of the relative width of the barrier, $\alpha$ (insets).
Upper inset: $y_{34}(\alpha)$. The variation in the interval
2.5$<\alpha <$3.0 is less than 10\% . 
Lower inset: linear dependence of ${\mathrm ln} (y_5)$ on $\alpha$. 
Note that $y_5(3)/y_5(2.5)\approx 5$. 
}
\label{Fig4}
\end{figure}

Fig.~\ref{Fig4} shows $n$-dependence of the ratios 
$y_n\equiv u_n/u_2=\tilde{u}_n/\tilde{u}_2$ 
(for $\varepsilon_F=10^3\varepsilon_3$, $T=0.00204\varepsilon_3$, and 
the relative width of the barrier $\alpha$ equal to 2.5). 
One can see that $y_3\gg 1$, i.e., $u_3\gg u_2$.
As we expected, absolute values of $y_4,y_5,y_6,$ are also large, 
but the signs alternate.
In agreement with our analytical conclusions 
further increase of $n$ gradually reduces $|y_n|$, 
and $y_n\to y_{\infty}\sim 1$ when $n\to \infty$.  

On the insets of Fig.~\ref{Fig4} we present
$\alpha$-dependencies of $y_n$ to make it evident that
although $u_{n>2}$  {\em is not exponentially small} as $u_2$ is,
it regains this smallness as $n\to \infty$.
Indeed, $y_{34}$ is almost a constant in the interval 2.5$<\alpha <$3.0
whereas $y_5$ increases with $\alpha$ by factor $\sim$5 in the same interval,
and $\ln (y_5)$ is a linear function of $\alpha$.

\begin{figure}[ht]
\epsfxsize=8cm
\centerline{\epsfbox{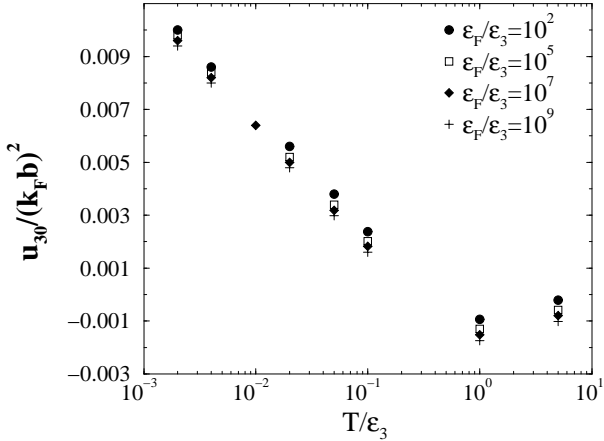}
}
\caption{ The dependence of the quantity
$\tilde{u}_{30}\equiv u_{30}(k_Fb)^{-2}$ 
on ${\mathrm ln}\left(T/\varepsilon_3\right)$
for different $\varepsilon_F$ shown in the legend.
}
\label{Fig5}
\end{figure}

We also computed the temperature dependence
of  $\tilde{u}_{30}\simeq \tilde{u}_{\infty}$
for several Fermi energies.
This dependence is presented on Fig.~\ref{Fig5} in semilogarithmic scale.
All four curves coincide , i.e., $\tilde{u}_{\infty}$
{\em does not depend on} $\varepsilon_F$. Moreover, as it was expected,
the logarithmic dependence, $u_{\infty}\propto \ln (T)+const$, persists
only as long as $T<\varepsilon_3$

One can see that the numerical simulations
unambiguously support the analytical conclusions. 
Taking into account the excited levels
{\em does not remove exponential smallness}
of the second order correction to the scattering
amplitude. A similar problem with the similar solution - 
contribution of the continuous spectrum above the barrier 
to the $\alpha$-decay rate - is
described in the book~\cite{Migdal}.

Returning to the Kondo temperature, Eq.~(\ref{TKcorrect}), 
we find that $T_K\leq 10^{-5}$ for 
the "typical parameters" Eq.~(\ref{typical}) and an optimistic 
estimate  $\varepsilon_3\approx 50K$. 
Therefore the Kondo model 
based on movable structural defects is hardly able to explain the
experiments, ~\cite{RLDB,MW,Pothier}.

Moreover, the estimate (\ref{TKcorrect}) excludes the very possibility
of the development of the strong coupling two-channel Kondo regime 
at arbitrary low temperatures:
it implies that the splitting of the two lowest levels of a TLS , 
$\Delta$, Eq.~(\ref{Delta}), always exceeds $T_K$. 
  Indeed,
$\Delta \geq \Delta_z \simeq \varepsilon_3 e^{-\eta} \gg v_z \varepsilon_3
\simeq\lambda(k_Fa)^2{\mathrm e}^{-\eta}\varepsilon_3$. 
Using the fact that the model is applicable only in the limit
 $v_z \ll v_x \ll 1$ we obtain
\begin{equation}
\frac{\Delta_z}{T_K} \gg
\left(\frac{4v_x}{v_z}\right)^\gamma \gg 1, \quad
\gamma = {\frac{1}{4v_x} -\frac{1}{2}}
\gg 1.
\label{hana}
\end{equation}
The same conclusion can be reached for any double-well potential model. 
Therefore {\em a movable defect weakly coupled
with electrons is unable to demonstrate 
the two channel Kondo non-Fermi liquid behavior.}~\cite{2loop}. 

We are grateful to  G. Zar{\'a}nd, A. Zawadowski, I. Smolyarenko 
and N. Wingreen for useful discussions. 
The work at Princeton University was supported by ARO MURI DAAG55-98-1-0270.
I.A. is a Packard Fellow.

\widetext

\begin{references}
\bibitem{2loop} The suppression of the tunneling splitting due to the
orthogonality catastrophe $\Delta_z \to \tilde{\Delta_z}\equiv 
\Delta_z (\epsilon_3/T_K){-v_x^2}$,
slightly changes index in Eq.~(\ref{hana}): $\gamma \to \tilde{\gamma}\equiv 
1/(4v_x)-1/2 - v_x$. One can see that $\tilde{\gamma}\approx \gamma \gg 1$, 
since $v_x\ll 1$. Therefore the inequality (\ref{hana}) remains valid 
even after taking into account the renormalization of $\Delta_z$.
\end{references}
\end{document}